\documentclass[12pt]{article}
\usepackage{amsxtra,amssymb,amsthm,amsmath,latexsym}

\theoremstyle{plain}

\newtheorem{theorem}{Theorem}[section]

\newtheorem{remark}{Remark}[section]
\numberwithin{equation}{section}

\def\ds{\displaystyle}
\def\nd{\noindent}

\def\oH{\buildrel\circ\over H}
\def\oH1{\buildrel\circ\over H\kern-.02in{}^1}
\def\qed{{\hfill $\Box$}}

\def\const{\hbox{\,const\,}}

\begin{document}
Jour. Math. Anal.Appl., 258, N1, (2001), 448-456.

\title{                  
Linear ill-posed problems and dynamical systems
   \thanks{key words: ill-posed problems, dynamical systems,
operator equations in Hilbert spaces     }
   \thanks{Math subject classification: 35R30, 47H17, 65J15 }
}

\author{
A.G. Ramm\\
 Mathematics Department, Kansas State University, \\
 Manhattan, KS 66506-2602, USA\\
ramm@math.ksu.edu\\
}

\date{}

\maketitle\thispagestyle{empty}

\begin{abstract}
A new approach to solving linear ill-posed problems is
proposed. The approach consists of solving a Cauchy problem
for a linear operator equation and proving that this problem has a global
solution whose limit at infinity solves the original linear equation.
\end{abstract}


\section{Introduction}
Let $A$ be a linear, bounded, 
injective operator on
a Hilbert
space $H$, and assume that $A^{-1}$ is unbounded
and that $\|A\|\leq \sqrt{m},$ where $m>0$ is a constant.
For example, $A$ may be a compact injective linear operator.
Consider the equation,
\begin{equation}\label{1.1}
   Au=f. \end{equation}
Assume that \eqref{1.1} is solvable, so that $f=Ay$
for a unique  $y\in H$. 
Problem \eqref{1.1}  is
ill-posed since $A^{-1}$ is unbounded. Equation \eqref{1.1} cannot be
solvable for all $f\in H$ because if $A$ is injective, linear, closed
and $R(A)=H$, then $A^{-1}$ must be bounded (by the Banach theorem).
Let $f_\delta$ be given, such that
\begin{equation} \label{1.2}
  \left\|f_\delta-f \right\|\leq \delta \end{equation}
Equation \eqref{1.1} with $f_\delta$ in place of $f$ may have no solution,
and if it has a solution $u_\delta$ then it may be that
$\left\|u-u_\delta\right\|$ is large, although $\delta>0$ is small.
There is a large literature on ill-posed problems since
they are important in applications. (See e.g.\cite{G},
\cite{E}).
In this paper a new approach 
 to solving linear ill-posed problems is proposed.
This approach consists of the following steps:

\vspace{.15in}
\nd \underbar{Step 1.} Solve the Cauchy problem:

\begin{equation} \label{1.3}
  \dot u=-[Bu+\varepsilon(t)u-F_\delta],
  \quad u(0)=u_0, \end{equation}
where

\[ \dot u:=\frac{du}{dt},\quad B:=A^\ast A, \quad
F_\delta:=A^\ast f_\delta, \quad ||F_\delta -F||\leq \delta \sqrt{m}, 
\quad F=By,
\]
\noindent
and

\begin{equation} \label{1.4}
\varepsilon(t)\in C^1 [0,\infty);\,\,
  \varepsilon(t)>0;
  \quad \varepsilon(t) \searrow 0 \hbox{\ as\ } t\to\infty;
  \quad\frac{|\dot\varepsilon(t)|}{\varepsilon^{\frac 52}(t)}\to 0
    \hbox{\  as\ } t\to \infty.
  \end{equation}
One has
$\left\|A^\ast(f_\delta-f)\right\|\leq \sqrt{m}\delta,$  
where we have used
the estimate  $||A||=||A^\ast||\leq \sqrt{m}.$

Examples of  functions $\varepsilon (t)$ satisfying  (1.4) can be 
constructed by the formula:
$$\varepsilon (t)=[c+ \int_0^t h(s)ds]^{- \frac 2 3},$$
where $c>0$ is a constant,  $h(s)>0$ is a continuous function
defined for all $s\geq 0$, such that $h(s) \to 0$ as $s \to \infty$
and $\int_0^\infty h(s)ds=\infty.$
One has
$\frac{|\dot\varepsilon(t)|}{\varepsilon^{\frac 52}(t)}=\frac
{2h(t)}3\to 0
    \hbox{\  as\ } t\to \infty.$
For example, $\varepsilon(t)=\frac 1{\log (t+2)}$ satisfies (1.4).
If 
$$h(t)=\frac 1{(2+t)\log (2+t)},$$
then
$$\varepsilon(t)=\frac 1{(1+ \log \log (2+t))^{\frac 23}}.$$
This $\varepsilon(t)$ yields nearly fastest decay of $h(t)$ allowed by
the restriction  $\int_0^\infty h(s)ds=\infty.$

\vspace{.15in}
\nd \underbar{Step 2.}
Calculate $u(t_\delta)$, where $t_\delta>0$ is a number
which is defined by formula \eqref{1.9}  below.

Then $t_\delta\to\infty$ as $\delta\to 0$ and satisfies the
inequality:
\begin{equation} \label{1.5}
  \left\|u(t_\delta)-y\right\|\leq 
\eta(\delta)\to 0 \hbox{\ as\ }\delta\to 0,
  \end{equation}
for a certain function $\eta(\delta)>0.$
If $\delta=0$, so that $F_\delta=By$, then Step 2 yields the relation
\begin{equation} \label{1.6}
  \lim_{t\to\infty} \|u(t)-y\|=0.
  \end{equation}

The foregoing approach is justified in Section 2.
Our basic results are formulated as follows.

\begin{theorem} 
Assume that  
equation \eqref{1.1} is uniquely solvable, \eqref{1.4} holds, and
$\delta=0$.
 Then  for any $u_0$, problem (1.3), with $F=By$ replacing
$F_\delta$, has a unique global solution and \eqref{1.6} holds.
\end{theorem}

By global solution we mean the solution defined for all $t>0$.

\begin{theorem} 
Assume that equation \eqref{1.1} is uniquely solvable,
\eqref{1.4} holds, and $\delta>0.$ 
Then
for any $u_0$ problem \eqref{1.3} has a unique global solution $u(t)$ and
there exists  a  $t_\delta \to \infty$ as $\delta \to 0$, such
that $||u(t_\delta)-y||\to 0$ as $\delta \to 0.$
The number $t_\delta$ is defined by formula \eqref{1.9}.
\end{theorem}

Let $y$ solve \eqref{1.1}. Then $By=F:=A^\ast f$ and $||B||\leq m$. If
\begin{equation}\label{1.7}
  \phi(\beta):=\phi(\beta, y):=\beta \left\| \int^{m}_0
   \frac{dE_\lambda y}{\lambda+\beta} \right\|, 
   \end{equation}
where $E_\lambda$ is the resolution of the identity of the selfadjoint
operator $B,$   $E_{\lambda -0}=E_\lambda,$
 $\beta(\delta)$  is the minimizer of the function
\begin{equation} \label{1.8}
  h(\beta,\delta):=\phi(\beta) + \frac
{\delta}{2\beta^{\frac 12}} 
  \end{equation}
 on
$(0,\infty),$  (see formula (2.20) and Remark 2.3 below),
and
\begin{equation} \label{1.9}
  \eta(\delta):=h(\beta(\delta),\delta),
  \quad \varepsilon(t_\delta)=\beta(\delta),
  \end{equation}
then $t_\delta\to\infty$ as $\delta\to 0$,
$\eta(\delta)\to 0$ as $\delta \to 0$, and
\begin{equation}\label{1.10}
  \lim_{\delta \to 0}\left\|u(t_\delta)-y\right\|=0.
  \end{equation}
Because $B$ is injective, zero is not an eigenvalue of $B,$
so, for any $y\in H,$ one has $||\int_0^s d E_{\lambda } y||\to 0$ 
 as 
$s\to 0.$ Therefore $\phi (\beta , y) \to 0$ as $\beta \to 0,$ 
for any fixed $y$.  From (2.15) (see below) one gets
\begin{equation} \label{1.11}
  \left\|u(t_\delta)-y\right\|<\eta(\delta) + g_{\delta}(t_\delta) \to 0   
\hbox{\ as\ }\delta \to 0,
  \end{equation}
where  $g_{\delta}(t)$ is given by the right-hand side of (2.12) with
$||f_\delta||$ replacing $||f||.$

\begin{remark}
Theorem 1.2 shows that solving the Cauchy problem \eqref{1.3} and
calculating
its solution at a suitable time $t_\delta$ yields a
stable solution to ill-posed problem \eqref{1.1} and
this stable approximate solution satisfies the error estimate
\eqref{1.11}.
\end{remark}

For nonlinear ill-posed problems a similar approach is proposed in \cite{R}.

\section{Proofs}

\subsection{Proof of Theorem 1.1}

We start with a simple, known fact:
if equation \eqref{1.1} is solvable, then it is equivalent to the equation
\begin{equation} \label{2.1}
  Bu=A^\ast f=By
  \end{equation}
Indeed, if $Ay=f$, then apply $A^\ast$ and get \eqref{2.1}.
Conversely, if \eqref{2.1} holds, then 
$(B(u-y),u-y)=\left\|A(u-y)\right\|^2=0$,
thus $Au=Ay$ and $u=y$, so \eqref{1.1} is solvable and its solution is the
solution to \eqref{2.1}.
Therefore we will study equation \eqref{2.1}. The operator $B=A^\ast A$ is
selfadjoint and nonnegative, that is, $(Bu,u)\geq 0.$ 
Let $E_\lambda$ be its resolution of the identity.

We make another observation: If \eqref{1.4} holds, then
\begin{equation} \label{2.2}
  \int^\infty_0 \varepsilon(t)dt=\infty.
  \end{equation}

Indeed, \eqref{1.4} implies
$$-\frac{\dot\varepsilon}{\varepsilon^2}\leq c,$$
where $c=\const>0$,
so 
$$\frac{d}{dt}\frac{1}{\varepsilon}\leq c,$$
$$\frac{1}{\varepsilon(t)}-\frac{1}{\varepsilon(0)}\leq ct,$$
$$\frac{1}{\varepsilon(t)}\leq c_0+ct,$$
$$c_0:=[\varepsilon(0)]^{-1}>0,$$
and
$$\varepsilon(t)\geq\frac{1}{c_0+ct}.$$
Formula  \eqref{2.2} follows from the foregoing inequality.

Consider the problem
\begin{equation}\label{2.3}
  Bw+\varepsilon(t)w-F=0,   \quad F:=A^\ast f=By.
  \end{equation}
Since $B\geq 0$ and $\varepsilon(t)>0$, the solution $w(t)$
of \eqref{2.3} exists, is unique and admits the estimate
\begin{equation}\label{2.4}
  \|w\| \leq \left\|(B+\varepsilon(t))^{-1} F\right\|
  \leq \frac{\|F\|}{\varepsilon(t)}\ .
  \end{equation}
If $F=A^*f,$ then (see Remark 2.3 below) one gets:
\begin{equation}\label{2.4'}
  \|w\| \leq \left\|(B+\varepsilon(t))^{-1} F\right\|=
  \left\|(B+\varepsilon(t))^{-1} A^*f\right\|
  \leq \frac{\|f\|}{2\varepsilon^{\frac 12}(t)}\ .
  \tag{2.4'}
  \end{equation}
Differentiate \eqref{2.3} with respect to $t$ (this is possible by
the implicit function theorem) and get
\begin{equation}\label{2.5}
  [B+\varepsilon(t)] \dot w=-\dot\varepsilon w,
  \quad \|\dot w\|
  \leq \frac{|\dot\varepsilon|}{\varepsilon} \|w\|
  \leq \frac{|\dot\varepsilon(t)|}{\varepsilon^2(t)} \|F\|,
  \end{equation}
where \eqref{2.4} was used.

Using \eqref{2.4'} yields:
\begin{equation}\label{2.5'}
\|\dot w\|
  \leq \frac{|\dot\varepsilon|}{\varepsilon} \|w\|
  \leq \frac{|\dot\varepsilon(t)|}{2\varepsilon^{\frac 32}(t)} \|f\|.
  \tag{2.5'}
  \end{equation}
Denote
\begin{equation}\label{2.6}
  z(t):=u(t)-w(t).
  \end{equation}
Subtract \eqref{2.3} from \eqref{1.3} (with $F$ in place of $F_\delta$)
and get
\begin{equation}\label{2.7}
  \dot z=-\dot w - [B+\varepsilon(t)]z,
  \quad z(0)=u_0-w(0).
  \end{equation}
Multiply \eqref{2.7} by $z(t)$ and get
\begin{equation}\label{2.8}
  (\dot z,z)=-(\dot w,z)-(Bz,z)-\varepsilon(t)(z,z).
  \end{equation}
Denote
\begin{equation}\label{2.9}
  \|z(t)\|:=g(t)
  \end{equation}
Then  the  inequality $(Bz,z)\geq 0$ and equation
  \eqref{2.8} imply:
\begin{equation}\label{2.10}
  g\dot g\leq \|\dot w\|g-\varepsilon(t)g^2.
  \end{equation}
Because $g\geq 0$, it follows from (2.10) and (2.5') that
\begin{equation}\label{2.11}
  \dot g\leq \|f\| \frac{|\dot\varepsilon(t)|}{2\varepsilon^{\frac
32}(t)}
  - \varepsilon(t)g(t),   \quad g(0)=\|u_0-w_0\|,
  \end{equation}
so
\begin{equation}\label{2.12}
  g(t)\leq e^{-\int^t_0 \varepsilon(s)ds}
  \left[ g(0)+\int^t_0 e^{\int^\tau_0\varepsilon(s)ds}
  \frac{|\dot\varepsilon(\tau)|}{2\varepsilon^{\frac 32}(\tau)}\ d\tau
\|f\| \right].
  \end{equation}

Assumption \eqref{1.4} (the last one in \eqref{1.4}) and \eqref{2.12} imply
(use L'Hospital's rule) that
\begin{equation}\label{2.13}
 \left\|u(t)-w(t)\right\| :=g(t)\to 0 \hbox{\ as\ } t\to +\infty.
 \end{equation}

The existence of the global solution to \eqref{1.3}
is obvious since equation \eqref{1.3}
is linear and the operator $B$ is bounded.

To prove \eqref{1.6} it  is sufficient to prove that
\begin{equation}\label{2.14}
  \|w(t)-y\|\to 0 \quad \hbox{\ as\ } t\to \infty.
  \end{equation}
Indeed, if \eqref{2.14} holds then \eqref{2.13} and \eqref{2.14} imply:
\begin{equation}\label{2.15}
  \|u(t)-y\| \leq \|u(t)-w(t)\| + \|w(t)-y\| \to 0
  \hbox{\ as\ }t\to \infty.
  \end{equation}

We now prove \eqref{2.14}. One has:
\begin{equation}\label{2.16}
  \|w(t)-y\| =
  \left\| \int^{m}_0 \frac{\lambda}{\lambda+\varepsilon(t)}dE_\lambda y
      -\int^{m}_0 dE_\lambda y \right\|
  =\left\| \int^{m}_0 \frac{\varepsilon(t)}{\lambda+\varepsilon(t)}
   dE_\lambda y\right\|.
   \end{equation}
Thus
\begin{equation}\label{2.17}
  \|w(t)-y\|=\phi(\varepsilon(t), y),
  \end{equation}
where $\phi(\varepsilon, y):=\phi(\varepsilon)$ is as defined in
\eqref{1.7}. Since $B$ is injective,
the point $\lambda=0$ is not an eigenvalue of $B$. Therefore 
\begin{equation}\label{2.18}
  \lim_{\varepsilon\to 0}\phi(\varepsilon)=0,
  \end{equation}
by the Lebesgue dominant convergence theorem.

Thus \eqref{2.14} follows and Theorem 1.1 is proved. \qed

\subsection{Proof of Theorem 1.2}

The proof is quite similar to the above, so we indicate only the new points.
Equation \eqref{2.3} is now replaced by the equation
\begin{equation}\label{2.19}
  Bw+\varepsilon(t)w-F_\delta=0.
  \end{equation}

Estimates \eqref{2.4}, \eqref{2.4'}, \eqref{2.5},
\eqref{2.5'} and \eqref{2.13} hold with $F_\delta$
and $f_\delta$ in
place of $F$ and $f,$ respectively.
The main new point is the estimate of $w(t)-y$:
\begin{align}\label{2.20}
  \|w(t)-y\|
  = & \left\| \int^{m}_0
     \frac{dE_\lambda F_\delta}{\lambda+\varepsilon(t)}
     - \int^{m}_0 dE_\lambda y \right\|     \notag\\
 = &
  \left\| \int^{m}_0
      \frac{dE_\lambda(F_\delta-F)}{\lambda+\varepsilon(t)} \right\|
      +\phi(\varepsilon(t))                      \notag \\               
 & \quad  \leq  \phi(\varepsilon(t))
  +\frac{\delta}{2\varepsilon^{\frac 12}(t)},
  \end{align}
where $||f-f_{\delta}||\leq \delta$ and estimate (2.4') was used.

If $\beta(\delta)$ is the minimizer of the function (1.8),
then
\begin{equation}\label{2.21}
  h(\beta(\delta),\delta):=\eta(\delta ) \to  0
  \,\, \hbox{\ as\ } \delta\to 0;
  \quad \beta(\delta) \to 0 \hbox{\ as\ }\,\, \delta\to 0.
  \end{equation}
The latter relation in (2.21) holds because $\phi(\beta)\to 0$ as
$\beta\to 0$.

Since $\varepsilon(t)\searrow 0$ as $t\to\infty$,
one can find the unique $t_\delta$ such that
\begin{equation}\label{2.22}
  \varepsilon(t_\delta)=\beta(\delta)\to 0\hbox{\ as\ }\delta\to 0.
  \end{equation}
Thus
\begin{equation}\label{2.23}
  \|w(t_\delta)-y\| \leq \eta(\delta)\to 0 \hbox{\ as\ } \delta\to 0.
  \end{equation}
The function  $\eta(\delta)=\eta(\delta, y)$ depends on $y$
because $\phi(\varepsilon)=\phi(\varepsilon, y)$ does 
(see formula \eqref{1.7}).

Combining \eqref{2.23}, \eqref{2.13} and \eqref{2.15} one gets
the conclusion of
Theorem 1.2. \qed

\begin{remark}

We also give a proof of \eqref{2.14} which does not use the spectral
theorem.

From \eqref{2.3} one gets
\begin{equation}\label{2.24}
  Bx+\varepsilon(t)x=-\varepsilon(t)y,  \quad x(t):=w(t)-y.
  \end{equation}
Thus $(Bx,x)+\varepsilon(x,x) = -\varepsilon(y,x)$.
Since $(Bx,x)\geq 0$ and $\varepsilon>0$, one gets
\begin{equation}\label{2.25}
  (x,x)\leq |(y,x)|, \quad \|x(t)\| \leq \|y\|=\const <\infty.
  \end{equation}

Bounded sets in $H$ are weakly compact. Therefore there exists a sequence
$t_n\to\infty$ such that
\begin{equation}\label{2.26}
  x_n:=x(t_n)\rightharpoonup x_\infty, \quad n\to\infty
  \end{equation}
where $\rightharpoonup$ stands for
the weak convergence. From \eqref{2.24} and \eqref{2.25}
it follows that
\begin{equation}\label{2.27}
  Bx_n\to 0, \quad n\to \infty.
  \end{equation}

A monotone hemicontinuous operator is weakly closed. This claim,
which we prove below, implies that \eqref{2.26} and \eqref{2.27}
yield $Bx_\infty=0$. because $B$ is injective,
$x_\infty=0$, that is, $x(t_n) \rightharpoonup 0$.
From \eqref{2.25} it follows that $\|x(t_n)\|\to 0$ as $n\to\infty$,
because $(y,x(t_n))\to 0$ as $n\to\infty$,
due to $x(t_n) \rightharpoonup 0$.
By the uniqueness of the limit, one concludes that
$\ds\lim_{t\to\infty} \|x(t)\|=0$,
which is \eqref{2.14}.

Let us now prove the claim.

 We wish to prove that $x_n \rightharpoonup x$ and
$Bx_n\to f$ imply $Bx=f$ provided that $B$ is monotone and hemicontinuous.
The monotonicity implies
$(Bx_n-B(x-\varepsilon p), x_n-x+\varepsilon p)\geq 0$
for all $\varepsilon>0$ and all $p\in H$.
Take $\varepsilon\to 0$ and use hemicontinuity of $B$ to get
$(f-Bx,p)\geq 0$ \quad $\forall p\in H$.
Take $p=f-Bx$ to obtain $Bx=f$, as claimed. \qed

The above argument uses standard properties of monotone  hemicontinuous
operators \cite{D}.
\end{remark}

\begin{remark} In \eqref{2.23}  $\eta (\delta)=O(\delta^{\frac 23})$
is independent of $y$ if $y$ runs through  a set
$ S_a:=S_{a,R}:=\{y: y=B^ah,\,\, ||h||\leq R \}$,  where
$R>0$ is an arbitrary 
 large fixed number and $a\geq 1.$
 If $0< a<1,$  then   $\eta (\delta)= O(\delta^{\frac {2a} {2a+1}}),$
as $\delta \to 0,$ and this estimate is uniform
with respect to $y \in  S_{a,R}$.

Consider, for example, the case $a\geq 1.$
If $y=B^a h$, then  $\phi(\varepsilon)$ in \eqref{2.20}
can be chosen for all
 $y \in  S_{a,R}$ simultaneously.  Using \eqref{1.7}, one gets:

\[ \phi(\varepsilon)= 
\varepsilon \sup_{||h|| \leq R}||\int_0^m \frac {\lambda^a}{\lambda +
\varepsilon} dE_{\lambda}h|| \leq 
\varepsilon m^{a-1} R,
\]
\noindent
where $a\geq 1$ and $\varepsilon$ is positive and small.
 For a fixed $\delta >0$ one finds the minimizer  
$\varepsilon (\delta)=O(\delta^{\frac 23})$ of the
function $\frac {\delta}{2\varepsilon^{\frac 12}}  + \varepsilon m^{a-1}R$
and the minimal value $\eta (\delta)$
 of this function is  $O(\delta^{\frac 23})$ .

If $B$ is compact, then
the condition $y\in S_a$ means that $y$ belongs to
a compactum which is the image
of a bounded set $||h||\leq R$ under the map $B^a$.

The case $0<a<1$ is left to the reader. 
It can be treated  by the method used above.
\end{remark}

\begin{remark} 
It can be  checked easily that
$$A(A^*A+\epsilon I)^{-1}=(AA^*+\epsilon I)^{-1}A.$$
This implies
$$||(B+\epsilon I)^{-1} A^*f||^2=((b+\epsilon I)^{-2}bf,f):=J,$$
where $B:=A^*A$ and $b:=AA^*\geq 0.$ 

Thus,
$$J=\int_0^m s(s+\epsilon)^{-2}d(e_sf,f)
\leq \frac 1 {4\epsilon}||f||^2,$$
where $e_s$ is the resolution of the identity corresponding to
the selfadjoint operator $b$.
Therefore one gets the following estimate:
$$||(B+\epsilon I)^{-1} A^*f||\leq \frac 1 {2\sqrt {\epsilon}}||f||.$$
 This estimate was used to  obtain estimates \eqref{2.4'} and \eqref{2.5'}.
 For
example, estimate \eqref{2.4} was replaced by the following one:

\begin{equation}
  ||(B+\epsilon I)^{-1}F||\leq \frac 1 {2 \epsilon^{\frac 12} 
}||f||,
  \tag{2.4'}
  \end{equation}
and \eqref{2.5} can be replaced by the estimate:

\begin{equation}
  \|\dot w\|
  \leq \frac{|\dot\varepsilon|}{\varepsilon} \|w\|
  \leq \frac{|\dot\varepsilon(t)|}{2\varepsilon^{3/2}(t)} \|f\|.
  \tag{2.5'}
  \end{equation}
\end{remark}

These estimates were used to improve the estimate for $\eta(\delta)$ in
the previous remark.

Estimate (2.4') was used by a suggestion of a referee. The
author thanks the referee for the suggestion.

In fact, one can prove a stronger estimate than (2.4'), namely
$||w||\leq ||y||$. Indeed, multiply (2.3) by $w-y$, use the
nonnegativity of $B$ and positivity of $\epsilon$ and get
$(w,w-y) \leq 0$. Thus $||w||^2 \leq ||w|| ||y||$, and the desired
inequality $||w||\leq ||y||$ follows. 

{\bf Appendix.}
Let us give an alternative proof of Theorem 1.2.
Let $u_{\delta}(t)$ solve (1.3), $u(t)$ solve (1.3)
with $F_{\delta}$ replaced by $F$, and  $u_{\delta}(t)$
 and  $u(t)$ satisfy the same initial condition. Denote 
$w_{\delta}:=u_{\delta}(t)-u(t)$ and let $||w_{\delta}||:=g_{\delta}(t)$.
One has: 

$$ 
  \dot w_{\delta}=-[Bw_{\delta}+\varepsilon(t)w_{\delta}-h_\delta],
  \quad  w_{\delta}(0)=0, 
$$
where $h_\delta:=F_{\delta}-F$, $||h_\delta||<\sqrt{m}\delta:=c\delta$.
Multiply the above equation by $ w_{\delta}$ in $H$, use 
the inequality $B\geq 0$  and get
$$
\dot g_{\delta}\leq -\epsilon(t) g_{\delta}+c\delta.
$$
Since $g_{\delta}(0)=0$, this implies:
$$
 g_{\delta}(t)\leq c\delta \exp[-\int_0^t \epsilon(s)ds]\int_0^t
\exp[\int_0^p \epsilon(s)ds]dp\leq c\frac {\delta}{\epsilon(t)}.
$$
Thus
$$
||u_{\delta}(t)-y||\leq ||u_{\delta}(t)-u(t)||+ ||u(t)-y||\leq
c\frac {\delta}{\epsilon(t)}+a(t), 
$$
where $a(t):=||u(t)-y||\to 0$ as  $t\to \infty.$
Define $t_{\delta}$ as the minimal minimizer of the 
following function of $t$
for a fixed $\delta>0$:
$$c\frac {\delta}{\epsilon(t)}+a(t)=min:=\mu(\delta).
$$
Since $a(t)\to 0$ and  $\epsilon(t)\to 0$ as  $t\to \infty,$
one concludes that the minimal minimizer $t_{\delta}\to \infty$
as $\delta \to 0$ and $\mu(\delta) \to 0$ as  $\delta \to 0$.
Theorem 1.2 is proved $\Box$.

\end{document}